Harnessing metastability for grain size control in multiprincipal element alloys during additive manufacturing

**Authors:** Akane Wakai[1], Jenniffer Bustillos[1], Noah Sargent[2], Jamesa Stokes[3], Wei Xiong[2], Timothy M. Smith[3], Atieh Moridi[1]*

**Affiliations:**

[1]Sibley School of Mechanical and Aerospace Engineering, Cornell University; Ithaca, NY, 14853, USA.

[2]Department of Mechanical Engineering and Materials Science, University of Pittsburgh; Pittsburgh, PA, 15260, USA.

[3]NASA Glenn Research Center; Cleveland, OH, 44135, USA.

*Corresponding author. Email: moridi@cornell.edu

**Abstract:**

Controlling microstructure in fusion-based metal additive manufacturing (AM) remains a challenge due to numerous parameters directly impacting solidification conditions. Multiprincipal element alloys (MPEAs) offer a vast compositional design space for microstructural engineering due to their chemical complexity and exceptional properties. Here, we establish a novel alloy design paradigm in MPEAs for AM using the FeMnCoCr system. By exploiting the decreasing phase stability with increasing Mn content, we achieve notable grain refinement and breakdown of columnar grain growth. We combine thermodynamic modeling, operando synchrotron X-ray diffraction, multiscale microstructural characterization, and mechanical testing to gain insight into the solidification physics and its ramifications on the resulting microstructure. This work paves way for tailoring grain sizes through targeted manipulation of phase stability, thereby advancing microstructure control in AM.

**Main:**

Multiprincipal element alloys (MPEAs), also known as high entropy alloys, are noted for their excellent mechanical and physical properties stemming from their unique chemical complexity. Unlike conventional alloys, MPEAs consist of multiple elements in high concentration. MPEAs comprising 3d transition metals have been widely explored due to their extraordinary strength and ductility at room (*1–3*) and cryogenic temperatures (*4*) owing to their low stacking fault energies and consequent additional deformation mechanisms (*5*). In additive manufacturing (AM), MPEAs offer a compelling avenue for the fabrication of high-performance components. Processing MPEAs via AM has led to improvements in properties such as strength (*6*, *7*) and corrosion resistance (*8*) compared to parts fabricated by conventional manufacturing methods such as casting. However, the AM processing of MPEAs presents challenges similar to conventional alloys, particularly in material systems that solidify along preferred crystal orientations (*9*). The high thermal gradients, rapid cooling rates, and partial remelting of previous layers can lead to the growth of columnar grains, resulting in highly textured microstructures. Advances in understanding of the composition-driven process-structure-property relationships in MPEAs will enable the strategic use of targeted compositions at specified locations to enhance the strength and performance of printed components (*10*).

Notable alloy design strategies for grain refinement in AM include combining pre-existing alloys such as titanium alloys, nickel-based superalloys, and stainless steels to create new materials (*11*, *12*), decorating the powder feedstock with nucleant particles (*13*), and adding alloying elements to increase constitutional supercooling ahead of the solidification front (*14*). In MPEAs, one effective method for changing the microstructure is to introduce a secondary phase by varying the composition of the alloy, such as adding aluminum, titanium, and niobium to MPEAs with 3d transition metals (15–20). Secondary phases such as body-centered cubic (bcc), Huesler-like ordered L2$_1$ phase, and Laves phase emerge to pin grain boundaries and break down epitaxial grain growth (21). However, these additional elements may lead to the formation of brittle intermetallics, which can result in significant cracking during solidification as they cannot accommodate the high residual stresses in AM. In addition, the underlying mechanism by which phase transformation leads to grain refinement remains elusive due to challenges with measuring and interpreting the spatiotemporal evolution of microstructures and defects after solidification has finished (22, 23). The current study aims to illuminate a new grain refinement mechanism in MPEAs by combining thermodynamic modeling, operando X-ray diffraction (XRD) studies, and multiscale microstructural evaluation to develop a comprehensive understanding of the intricate microstructural changes occurring during the AM process. We demonstrate an alloy design concept and adaptation for AM using a Fe$_{80-x}$Mn$_x$Co$_{10}$Cr$_{10}$ (at. %, x = 40, 45, and 50) MPEAs. These samples will henceforth be referred to by their Mn content, namely, Mn40, Mn45, and Mn50. Here, we increase the Mn content to destabilize the primary fcc phase during solidification to enable microstructure control in a direct energy deposition (DED) system where the microstructure generally tends to be more columnar due to its lower scanning speeds compared to powder-bed fusion (PBF) (24). FeMnCoCr enables us to systematically establish the complex relationship of composition, phase stability, process, microstructure, and properties to achieve a crack-free, refined microstructure in AM.

Solidification pathways

Phase stability is predominantly influenced by composition, as each element impacts the overall solidification behavior. In the FeMnCoCr system, the Scheil-Gulliver model (25) predicts the stable phase to be an fcc phase (γ), with a metastable bcc (δ) emerging during the initial stages of solidification (Fig. 1A). Operando XRD studies were conducted at Cornell High Energy Synchrotron Source (CHESS) to reveal the solidification pathways of each composition under the AM condition simulating a DED environment (26, 27). Two detectors, a CdTe Eiger 500k area detector and a far-field GE 41-RT area detector, were placed to capture portions of the diffraction cones at azimuthal angles ($\eta$) of 172.4° ≤ $\eta$ ≤ -172.3° and -90.9° ≤ $\eta$ ≤ 91.8°, respectively (Fig. S1). The Eiger detector captured XRD data at 100 Hz to give insight into the fast evolution of diffraction patterns during the AM process, and the GE detector captured data over a larger portion of the cone at a frequency of 4 Hz.

Time-resolved, azimuthal integration plots from the Eiger data detect three peaks associated with fcc at the beginning of the experiment for all compositions (Fig. 1B). As the materials melt, the peaks' 2θ values decrease, corresponding to a rise in temperature (thermal expansion). The peaks then shift back to reflect the temperature decrease as the melt pool solidifies and cools down. The original powder peaks in Mn40 and Mn45 remain in the captured data even during melting (between t ≈ 3 s and t ≈ 4.5 s) due to diffraction from the residual powder surrounding the melt pool and deposited bead. These powder peaks during melting are ignored for analysis and interpretation. In the Mn50 sample, an additional peak appears at t = 3.68 s before the fcc peak appears at t = 3.72 s, only to disappear several milliseconds later, as shown in the inset of Fig. 1b-Mn50. This peak corresponds to a metastable bcc phase, confirmed with additional peaks observed on the GE detector annotated in blue (middle frame during solidification at t = 3.75 s in Fig. S2). Unlike all other peaks, which increase in 2θ to reflect cooling, this bcc peak shows a decrease in the 2θ value even though the laser has passed and melting has already occurred (i.e., no more external heating). This peak shift reflects a lattice parameter expansion whose possible causes and implications will be explored later in the Discussion Section. Lastly, there are additional rings corresponding to the tetragonal $MnO_2$ and $Mn_2O_3$ in all compositions throughout the process (Fig. S2).

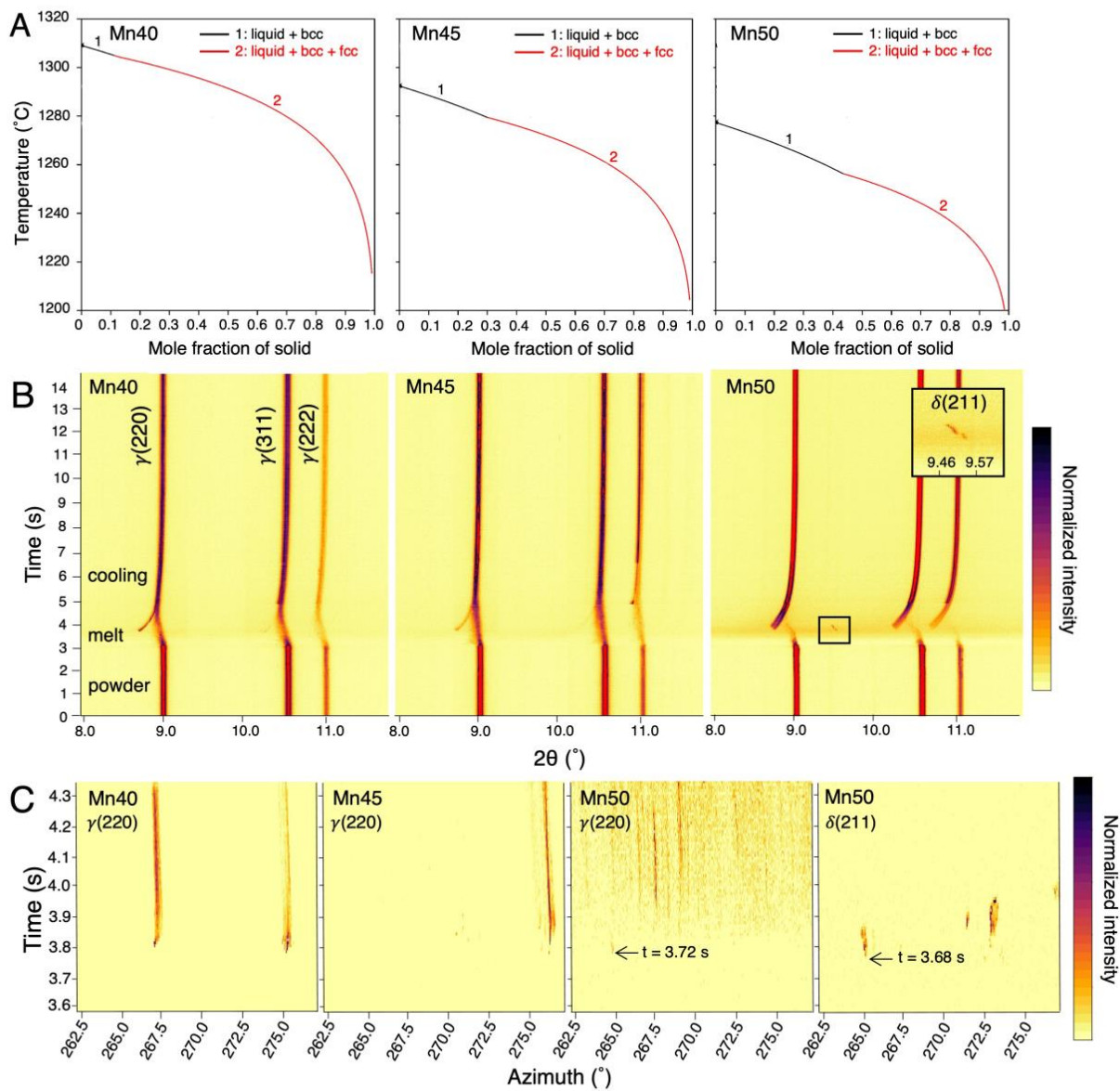

**Fig. 1. Solidification simulation and operando synchrotron XRD experimental results.** (A) Solidification pathway simulation of Mn40, Mn45, and Mn50 via the Scheil-Gulliver model using the Thermo-Calc TCHEA6 database (*28*). (B) 2θ *vs.* time plots for operando XRD. (C) Azimuth *vs.* time plots during melting and solidification of the γ(220) and δ(211) peaks for sample Mn50. This plot is representative of other azimuths.

For a more comprehensive understanding of the solidification behavior, we examine the raw datasets for information on the evolution of diffracted during the AM process. Azimuth *vs.* time plots shown in Fig. 1C illustrate the progression of diffraction spots during solidification in the $(220)_\gamma$ peak and $(211)_\delta$ peak for Mn50. Unlike the discrete spots in $(211)_\delta$, the fcc spots are more uniformly dispersed, a contrast that becomes pronounced hwen to the other two compositions. The data for Mn40 and Mn45 indicates limited grains aligning with the diffraction condition (267.5° and 274.9° in Mn40 and those at 276° in Mn45). Conversely, Mn50 exhibits peaks with lower intensities (only 4.8% of the maximum intensity of Mn40 and 38.5% of Mn45) and spread out more evenly, suggesting a larger number of grains satisfying the Bragg's condition across diverse crystallographic orientations. This distribution implies a smaller average diffracted volume, likely due to reduced grain sizes, as indicated by weaker signal per peak.

Multiscale microstructural evaluation

The microstructural analysis of the single-track beads obtained at CHESS highlight a remarkable trend, as shown in the backscatter electron (BSE) micrographs (Fig. 2A-C). Notably, Mn50 exhibits a significant grain size reduction of over 70% compared to the other compositions, with average grain sizes of 20.0±14.8 µm, 26.0±21.4 µm, and 5.3±3.8 µm for Mn40, Mn45, and Mn50, respectively (Fig. 2d-f), without changing the process parameters. The directional grain growth towards the build direction in Mn40 and Mn45 is interrupted in Mn50, leaving a more equiaxed microstructure. Electron backscatter diffraction (EBSD) confirms over 99% fcc in all three compositions at these pixel sizes and view fields (Fig. 2G-I). However, BSE imaging unveils the presence of a small fraction of a secondary phase in Mn50 scattered near the edge of the bead. These islands are indexed via EBSD as a tetragonal σ phase, as shown in the inset of Fig. 2C. The σ phase is typically found at grain boundaries and is 2.7±1.9 µm in size. Synchrotron XRD data does not capture any σ phase likely because the overall intensities from the σ phase tend to be much weaker than those from the fcc and bcc peaks due to its low-symmetry crystal structure and low multiplicity factor *(29)*, which further increases the difficulty of detecting signal from the σ phase. The low volume fraction of σ phase and peak overlap with manganese oxides further complicate its detection. Finally, the σ phase was observed to occur near the bottom of the bead and may not have been within the X-ray window.

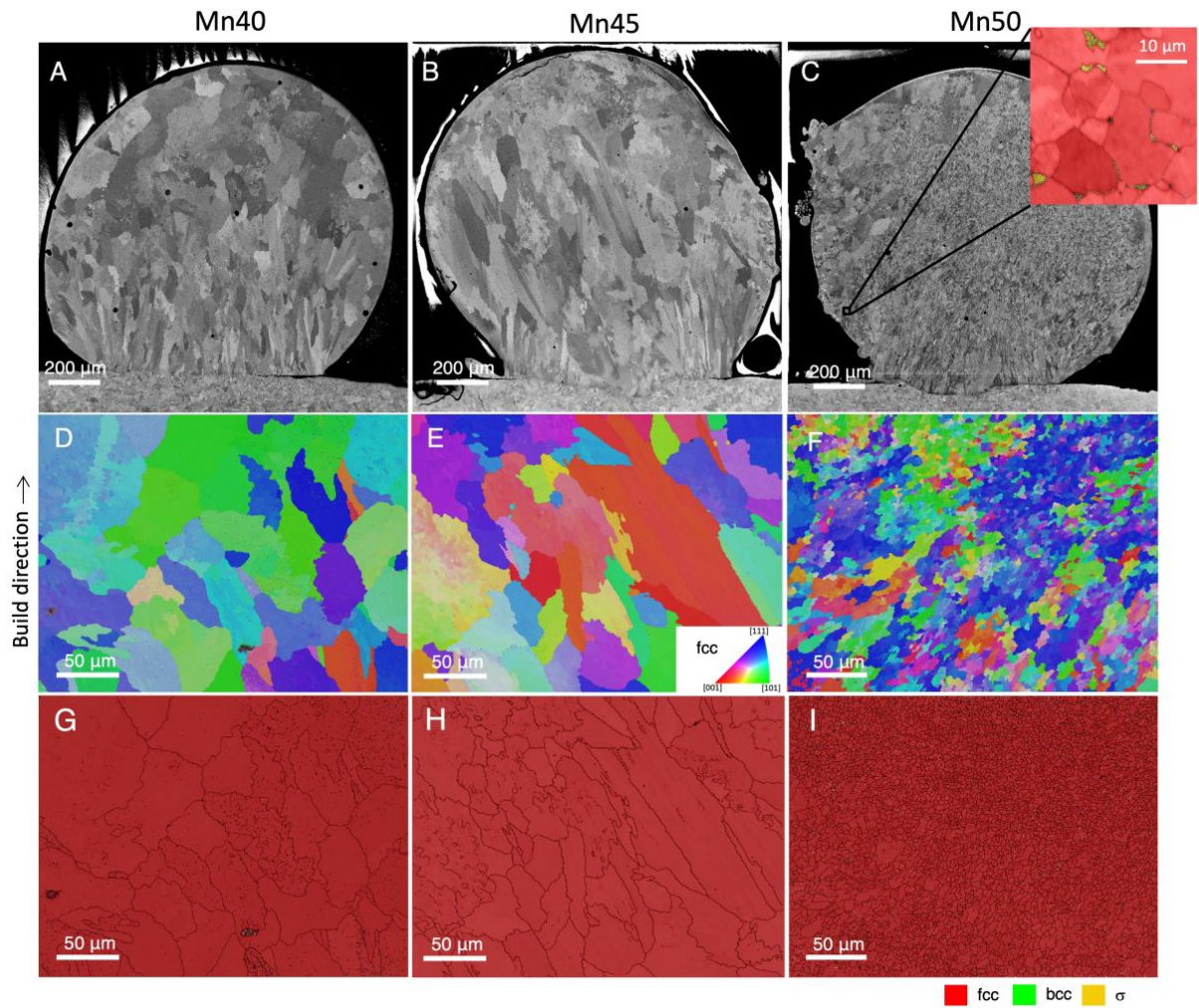

**Fig. 2. Scanning electron microscopy of single-track FeMnCoCr MPEAs.** (A-C) BSE micrographs of cross sections of single tracks whose operando XRD data were obtained for Mn40, Mn45, and Mn50, respectively. Inset in (C) shows the σ phase (yellow) forming along grain boundaries in Mn50 indexed via EBSD. (D-F) Inverse pole figures (IPFs) of Mn40, Mn45, and Mn50, respectively. (G-I) Phase maps of Mn40, Mn45, and Mn50, respectively. Black lines depict grain boundaries.

Printing the three compositions in multiple layers on a FormAlloy X2 DED system yields a microstructural trend consistent with the single-track beads, as revealed by EBSD shown in Fig. 3A-C. IPF maps depict a columnar grain growth oriented parallel to the build direction in Mn40 and Mn45 shown in Fig. 3A and 3B, respectively. Such large columnar grains are a direct result of the partial remelting of previous layers often present in as-printed conditions. By contrast, continuous grain growth across layer boundaries is interrupted at the melt pool boundary in Mn50 (Fig. 3C). Moreover, significant grain refinement accompanies the breakdown of epitaxial grain growth, with an average grain size of 41.6±27.8 µm in Mn50 compared to an average of 191.6±140.8 µm and 321.4±251.7 µm for Mn40 and Mn45, respectively. Pole figures shown below Fig. 3C indicate a reduction in texture in Mn50 compared to Mn40 and Mn45, with maximum multiples of a uniform density values decreasing from 3.70 and 3.37 to 2.10 in Mn40, Mn45, and Mn50, respectively.

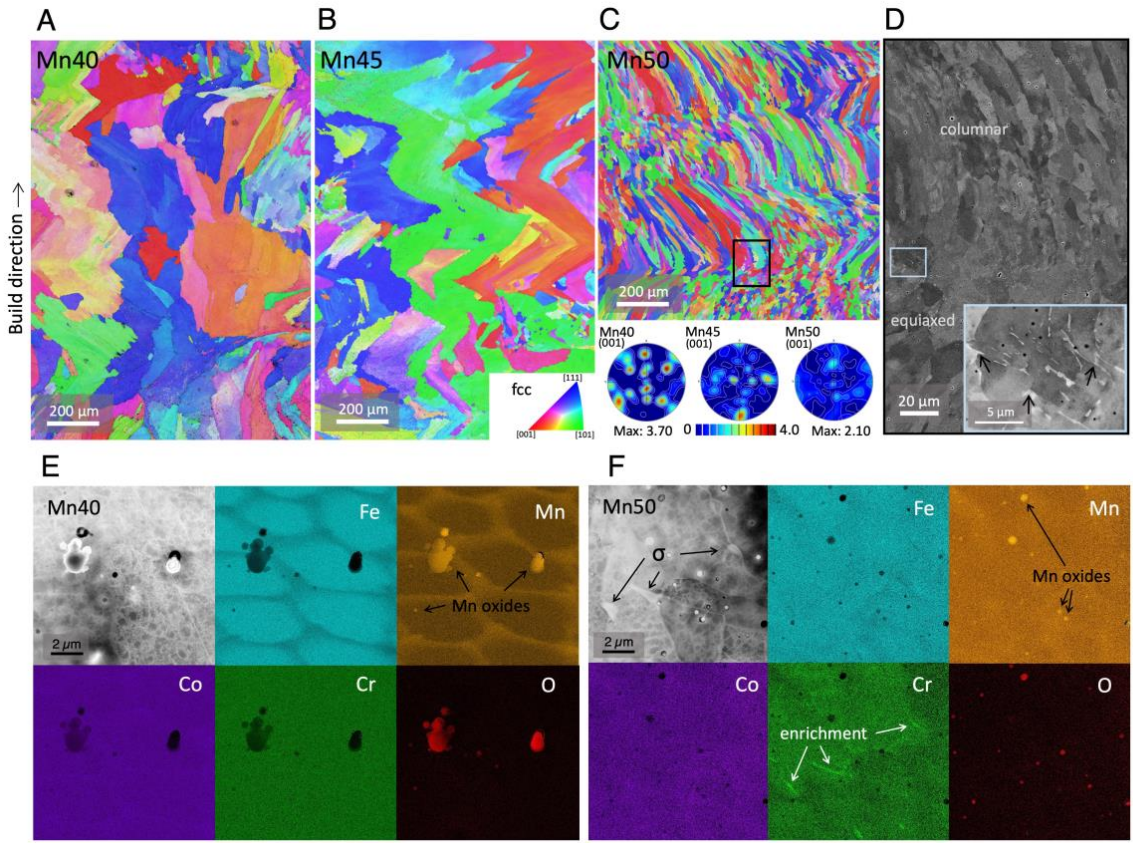

**Fig. 3. Electron microscopy of as-printed FeMnCoCr MPEAs.** (A-C) IPFs of Mn40, Mn45, and Mn50, respectively. Pole figures of IPFs in the (001) direction are shown below. (D) BSE image of columnar-to-equiaxed transition showing σ-phase precipitation. (E,F) STEM and EDS images of Mn40 and Mn50, respectively. The electron beam is parallel to the [110] zone axis of the matrix in both STEM images presented. High density of dislocations is seen in both materials. Secondary phase appears brighter than the fcc matrix in Mn50. Cr segregation (white arrows) correspond to interfaces between fcc and sigma phase.

The presence of σ islands is predominantly observed at the columnar-to-equiaxed transition as indicated by the black box in Fig. 3C. A BSE image of the boxed area shows a combination of the σ phase (bright spots) and the fcc matrix (dark regions). Prominent chromium enrichment is found at interfaces between the σ phase and fcc, as indicated by white arrows (Fig. 3F). Most σ islands appear at grain boundaries, as shown in Fig. 2C and 3D. Through scanning transmission electron microscopy (STEM), we observe a significant accumulation of dislocations in the fcc matrix in both Mn40 and Mn50 (Fig. 3E and 3F, respectively), which is typical of AM parts due to the residual stress from fast cooling and thermal cycling *(30)*.

Chemical composition changes between the feedstock powders and the as-printed samples are marginal (Table S1). STEM-energy dispersive x-ray spectroscopy (EDS) visualizes the depletion of Fe and Cr and the enrichment of Mn and Co in cellular structures more prominently in Mn40 than in Mn50 (Fig. 3E,F). This phenomenon is commonly observed in as-printed austenitic stainless steels like 304L and 316L *(31)* as well as as-printed MPEAs *(32)*. STEM-EDS also depicts the presence of manganese oxides scattered across both samples (Fig. 3E,F), which has also been observed in other as-printed MPEAs, such as the CoCrFeMnNi *(33)*. The oxides in Mn40 are significantly larger than those in Mn50, which corresponds with the higher intensities of signal from the oxides in the operando XRD data in Mn40 shown in Fig. S2. The initial oxygen composition of the feedstock (likely existing as an oxide film on the surface of the feedstock powder or oxide inclusions within the feedstock) may provide the oxygen to form the fine inclusions. Since the XRD signal from manganese oxides is more prominent in Mn40 even before the AM process, it is possible that the oxygen content in the powder is higher in Mn40 and resulted in larger oxide inclusions. Alternately, any oxygen in the atmosphere may be picked up during the print processes. These oxide inclusions nucleate and grow when the solubility of oxygen dissolved in the molten pool decreases (*31*).

Mechanical properties

Tensile testing was conducted to assess the mechanical performance of the as-printed MPEAs in a direction perpendicular to the build direction (Fig. 4). The yield strengths exhibited an upward trend, rising from 372.7±10.8 MPa, 378.2±11.6 MPa, to 411.9±18.3 MPa in Mn40, Mn45, and Mn50, respectively. There is a slight decrease in ductility with increasing Mn content (28.9±4.1%, 27.9±3.1%, and 26.8±1.0%, respectively). The grain refinement achieved in Mn50 can account for the observed strength enhancement in accordance with the Hall-Petch relationship. In comparison to the stress-strain curve of as-cast $Fe_{37}Mn_{45}Co_9Cr_9$ (*1*), which most closely matches the composition of Mn45 in this study, the as-printed material exhibited substantial strength improvements. The rapid solidification from AM induces a high density of dislocations, thereby contributing to the material's strengthened state (*19*). Ongoing investigations aim to provide a comprehensive understanding of the mechanisms influencing the mechanical properties of the as-printed MPEAs in this study.

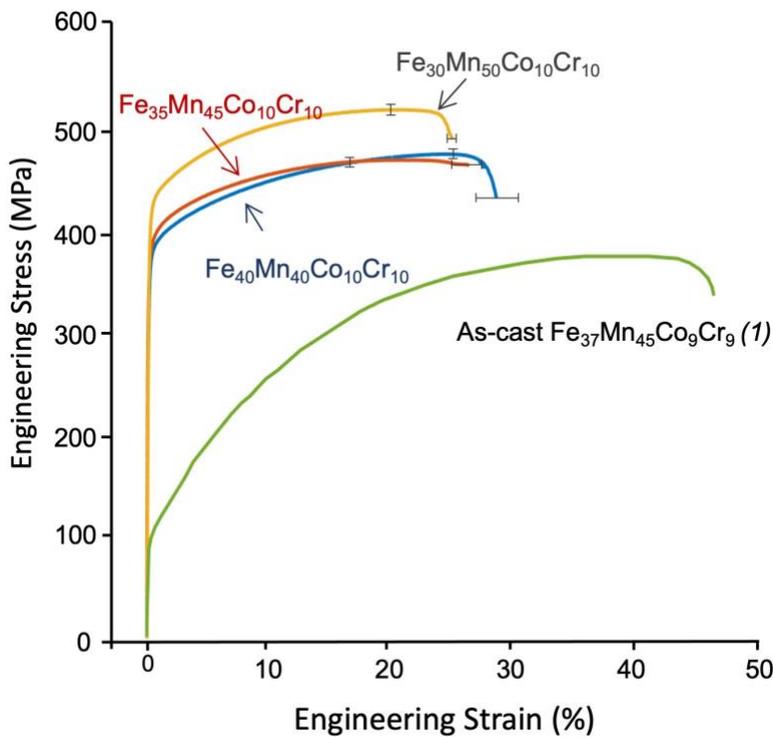

**Fig. 4. Stress vs strain curve of as-printed and as-cast (*1*) FeMnCoCr MPEAs.**

Discussion

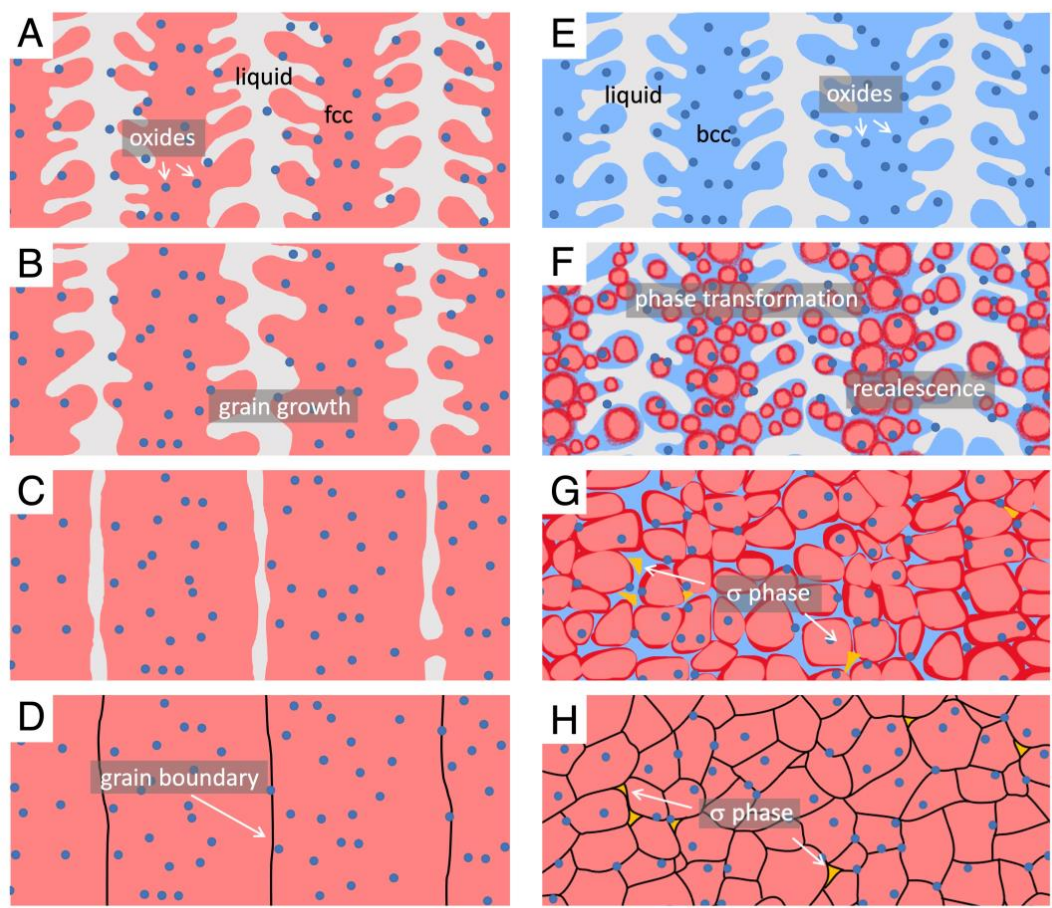

**Fig. 5. Proposed solidification mechanism.** (A-D) growth mechanism of Mn40 and Mn45. (E-H) Solidification and transformation mechanism of Mn50. Intermediate phase may act as a grain refiner through transformation to inhibit grain growth and coalescence.

A compelling correlation emerges between the destabilization of the fcc phase and grain refinement. Mn40 and Mn45, with higher fcc phase stability, directly solidify into fcc along with manganese oxides (Fig. 5A-D). Conversely, Mn50 follows a more complex solidification pathway, initially solidifying into bcc (Fig. 5E) with large, uniquely oriented grains based on the spotty diffraction pattern in azimuth vs. time plot (Fig. 1C). Transformation to fcc grains occurs from the bcc phase (Fig. 5F), marked by a sudden uniform diffraction signal after bcc signals. A similar mechanism is observed to break down grains in Fe-C via time-resolved X-ray imaging, where a massive δ-to-γ transformation results in the fragmentation of grains *(34)*. In addition, we hypothesize that the phase transformation may cause recalescence – the release of latent heat due to phase transformation – which may result in grain remelting to further enhance grain refinement. As mentioned previously, a peak shift towards a lower 2θ value is observed in the bcc peak, which signifies an increase in the lattice parameter. Multiple possible factors can contribute to this peak shift such as mechanical strain (due to a macrostress) *(35)*, compositional change (e.g. doping *(36)*), and a rise in temperature *(37)*, and it is difficult to decouple each one from the other. Although we are unable to completely rule out mechanical and chemical contributions to this observation, thermal effects are currently hypothesized to be a major contributor to the peak shift. In a first approximation, we assume that thermal effects are most dominant immediately following solidification when the cooling rates are the highest at ~1000 K/s (i.e. assume no contributions from mechanical and chemical effects), the maximum possible temperature rise detected by the synchrotron XRD would be 107.0±16.8°C. Previous studies have shown that δ-dendrite fragmentation can occur when dendrite arms remelt after recalescence during solidification from an undercooled melt (*38, 39*), and remelting has led to grain refinement for an undercooling between 60°C and 100°C in Ni-Cu systems (*40*). Given that the estimated temperature rise is of a similar magnitude to the recalescence-grain refinement phenomenon, it is plausible that the increasing temperature in the bcc phase could have also remelted the dendrite arms to cause grain refinement as illustrated in Fig. 5F and 5G. Further systematic investigations are needed to isolate effects from chemical, mechanical, and thermal effects on the resulting XRD data to validate this hypothesis.

The solidification pathway of the σ phase cannot be inferred from the operando XRD data because its signal could not be detected. We hypothesize that the intermetallic phase nucleates and grows from the intermediate bcc phase as often seen in duplex stainless steels. The bcc crystal structure is known to transform to the σ phase more quickly than fcc into σ due to factors such as its less densely packed crystal structure allowing for faster diffusion (*41*), incoherent crystallography between the σ phase and fcc (*42*), and Cr acting as stabilizers to both bcc and σ phases (*43*). While the σ phase can precipitate from the fcc phase in some 3d transition metal MPEAs, it usually occurs extensive heat treatment (*44*) or severe plastic deformation and recrystallization through annealing (*45*). In a reported case of as-printed PBF CoCrFeMnNi, a small amount of the σ phase is attributed to the small grain size and high dislocation density, which promote diffusion and are influenced by thermal cycling in AM (*46*). In this study, since the σ phase is generally stable between 600 and 1000 °C (*42, 47*), it is inferred that the bcc-to-σ transformation occurs in solid-state (Fig. 5H). The mechanism and kinetics of the intermetallic phase formation in this MPEA systems are subject to further investigation.

Lastly, we note the limitations of experimentation and simulations conducted for this study. The operando XRD data suggests bcc phase formation only in one composition, while the Scheil-Gulliver model predicts a bcc-to-fcc solidification pathway for all. Current analysis based on a single projection XRD without sample rotation limits the number of grains aligned in the diffraction condition. In addition, it is possible that bcc may form in Mn40 and/or Mn45 that goes undetected by the detectors if (1) the timescale of bcc formation and transformation in Mn40/Mn45 is faster than the temporal resolution of the current study and (2) the signal from bcc phase is too low for detection. Setup modifications such as integrating multiple detectors or increasing beam flux is necessary for further probing the fast evolution of microstructures and defects in AM. Moreover, the Scheil-Gulliver model does not consider solidification rates, which can greatly vary in AM systems (between $10^3$ K/s-$10^5$ K/s in DED and $10^4$ K/s-$10^6$ K/s in PBF systems (*24*)). Since solidification rates play a crucial role in determining the solidification pathway, relying solely on the Scheil-Gulliver solidification simulation may not always provide an accurate guideline for selecting metastable MPEAs suitable for all AM techniques. However, the trend found through Scheil-Gulliver – where the bcc formation increases from Mn40, Mn45, to Mn50 – remains valid. In light of these findings, the Scheil-Gulliver model can guide the initial composition selection, pending further experimental validation to accurately predict and design MPEAs for AM applications.

Phase stability engineering is demonstrated as a viable methodology for designing alloys specifically for AM by harnessing its unique processing conditions. The combined operando synchrotron XRD, thermodynamic modeling, microstructural analysis, and mechanical testing elucidates the composition, process, microstructure, and properties relationship. This methodology is applicable to other MPEA systems with an intermediate phase, but screening compositions via thermodynamic simulation is critical for an effective design approach to AM of MPEAs. These MPEAs exploit the intrinsic rapid solidification through phase transformations of metastable phases that break up continuous grain growth across layer boundaries. These findings set the stage for theory-guided exploration in the extensive compositional space of MPEAs for AM.

**Materials and Methods:**

Solidification Simulation

Scheil-Gulliver simulations were performed using Thermo-Calc software version 2023a with the TCHEA6 thermodynamic database to predict the solidification behavior of each alloy composition . The Scheil model leverages thermodynamic data to predict the phase evolution and corresponding composition variations during solidification. The model applied assumes homogeneity in the liquid composition, no diffusion in the solid phase(s), and equilibrium at the solid-liquid interface. Considering the Marangoni convection in AM that effectively homogenizes the melt pool(*48*), the assumptions made in the model are justified. Previous studies in the literature have shown agreement between simulations and experimental results(*49*).

Operando Synchrotron X-ray Diffraction

Synchrotron X-ray diffraction studies were conducted at Cornell High Energy Synchrotron Source (CHESS). A custom printer was integrated at the Forming and Shaping Technology ID3A (FAST) beamline(*26*). A high-energy, monochromatic X-ray beam with a wavelength of 0.2022 Å and energy of 61.322 keV was used in transmission mode with a square cross-section of 0.750 mm x 0.750 mm. A CdTe Eiger 500k area detector with 512 x 1024 pixels captured diffraction patterns at a frame rate of 100 Hz and covered azimuthal angles (η) between -172.3° and 172.4° and diffraction angles (2θ) between 7.5° and 11.2°. The detector-to-sample distance was 899 mm, which was calibrated using a $CeO_2$ reference powder. Additionally, a GE 41-RT area detector with 2048 x 2048 pixels was used to capture azimuthal angles between -90.9° and 91.8° and diffraction angles between 0.6° and 15.4° at a frame rate of 4 Hz.

Powder from each composition was rastered with a 500W continuous wave multi-mode laser from IPG Photonics at a laser power of 200W, scanning velocity of 4.5 mm/s, and layer height of 2 mm. These parameters were chosen to achieve a stable bead with the constraint of the scanning speed, which was the maximum velocity attainable on the Huber stage available at CHESS. Azimuthal integration and azimuth vs time plots of the Debye-Scherrer diffraction patterns from the Eiger detector was performed using GSAS-II(*50*). For azimuth vs time plots, 2θ values from 8.7° to 9.1° and 7.6° to 7.9° were binned and integrated for the γ(220) peaks and δ(211) peak in Mn50, respectively.

The calculation of temperature rise in the bcc phase was done by first fitting each bcc diffraction spot to extract the 2θ values from the Eiger detector on GSAS-II using the Pseudo-Voigt model. The corresponding d-spacings were calculated using the Bragg's Law, $n\lambda = 2d\sin(\theta)$. Together with the coefficient of thermal expansion (CTE), the change in temperature ($\Delta T$) was acquired using the following equation(*51*):

$$\Delta T = \frac{d_f - d_0}{d_0 * CTE}$$

Where $d_0$ and $d_f$ are the initial and final d-spacings, respectively. The reported uncertainty is taken as the standard deviation of the changes in temperature for all diffraction spots.

Direct Energy Deposition

FeMnCoCr spherical powders were gas-atomized (Arcast, Oxford, ME) and sieved to achieve particle diameters ranging between 15 and 45 µm. Printing was performed on a FormAlloy X2 DED system equipped with an IPG Photonics Nd:YAG continuous wave fiber laser with 500W maximum power at a spot size of 1.2 mm. The build chamber was purged with argon to reduce the oxygen level to below 100 ppm. Blocks of 26 mm x 10 mm x 6 mm (WxLxH) were printed on a 304L stainless steel substrate with a rectilinear infill scan strategy with hatch spacing of 0.6 mm and an angle offset of +67˚ and -67˚ in alternating layers with a 0.2 mm layer height. All compositions were printed under the same nominal processing conditions (laser power of 250W, scanning speed of 800 mm/min, powder feed rate of 0.5 rpm).

Scanning Electron Microscopy (SEM)

Single beads from Cornell High Energy Synchrotron Source (CHESS, Ithaca, NY) and as-printed samples were cross sectioned using a high-speed diamond saw parallel to the build direction and polished down to 0.08 µm colloidal silica for compositional and microstructural evaluation. Imaging and electron dispersive spectroscopy (EDS) were conducted on a Tescan Mira3 field-emission scanning electron microscope (FE-SEM) equipped with a backscattering detector. Electron backscatter diffraction (EBSD) was acquired on a QUANTAX EBSD for grain morphology and phase makeup evaluation (Bruker, Billerica MA). The field of view of 300 µm and pixel size of 0.5 µm were used for all single bead EBSD measurements. The field of view of 1000 µm and pixel size of 0.6 µm were used for multilayer Mn50, and a wider view field of 1500 µm and a pixel size of 1.2 µm were used for multilayer Mn40 and Mn45 to capture the larger grains. The measurements were processed using ATEX (Metz, France).

Transmission Electron Microscopy (TEM)

Scanning transmission electron microscope (STEM) disc samples (diameter 3 mm) were extracted via wire-electron discharge machining from Mn40 and Mn50. STEM discs were thinned manually to 100 µm using SiC polishing paper. The STEM discs were then electropolished with a 90% methanol and 10% perchloric acid at -40 ˚C and 12V using a Struers twin-jet polisher. STEM-EDS was conducted on a FEI Talos at 200 kV using a Super-X energy dispersive X-ray spectroscopy detector and processed using Thermo Fisher Scientific Velox software.

Coefficient of thermal expansion

The procedures to determine the coefficient of thermal expansion (CTEs) have been described in detail in a previous publication(*52*). In brief, the CTE was determined via *in situ* diffraction in an Empyrean diffractometer (Malvern PANalytical) using cobalt radiation (Co Kα = 1.789 Å) and a hot stage (Anton Paar HTK 2000N). National Institutes of Standards Technology (NIST) standard reference material (SRM) aluminum oxide ($Al_2O_3$) powder was used as a calibration standard dispersed onto a sacrificial piece of tantalum foil on the tantalum heating strip to determine the temperature set points. The measured values for $Al_2O_3$ were compared to the recommended values for the thermal expansion ($\Delta L/L_0$) as a function of temperature *T* (*53*), as

$$(a_0)\frac{\Delta L}{L_0} = -0.176 + 5.431 \times 10^{-4}T + 2.150 \times 10^{-7}T^2 - 2.810 \times 10^{-11}T^3$$

$$(c_0)\frac{\Delta L}{L_0} = -0.192 + 5.927 \times 10^{-4}T + 2.142 \times 10^{-7}T^2 - 2.207 \times 10^{-11}T^3$$

where temperature is measured in Kelvin. Equations (1) and (2) are then inverted, and the measured values of the average change in lattice parameters are used in the inverted equation to find the actual temperature of the heating strip. Once temperature set points were determined, the sample powders were dispersed onto tantalum foil on the heating strip. Scans were taken over a 2θ range of 40° to 130° at temperatures of 25°C, 200°C, 400°C, and 600°C. All scans were analyzed in TOPAS (54) using the LeBail method (55) to estimate the lattice parameters of the materials at each temperature. The average bulk CTE is approximated as 1/3 of the unit cell volume expansion coefficient (i.e., $\Delta V$ vs. $\Delta T$ divided by 3). The value of the coefficient of thermal expansion in Mn50 was determined to be $18.38 \times 10^{-6}$ K$^{-1}$.

Mechanical Characterization

The printed samples were machined perpendicular to the build direction via wire-EDM into micro-tensile specimens with a gauge length of 8 mm, width of 2 mm, and thickness of 0.7 mm. The specimens were ground down to 0.6 mm thickness to remove the oxide layer on both sides. The evaluation of the mechanical properties of the samples was conducted under tensile loads using a Deben MT 2000 micro-tensile stage equipped with a 2 kN load cell (Deben UK Ltd, Suffolk, UK). Tensile experiments were performed in displacement control mode at an average strain rate of $2.0 \times 10^{-3}$ min$^{-1}$. Non-contact, real-time evolution of strains was captured by a digital image correlation software (GOM, Braunschweig, Germany) from the recorded tensile displacements. Three tensile samples for each composition were tested to confirm reproducibility, and the average values and standard deviation are reported. The yield strength was determined with the 0.2% offset plastic strain method.

Cooling rates

The cooling rates of each composition were calculated based on the change in 2θ values during the operando X-ray diffraction experiments. The lattice parameter ($a$) was calculated for the γ(220) peak using Bragg's law:

$$a = \frac{\lambda\sqrt{h^2 + k^2 + l^2}}{2\sin(\theta)}$$

Where $\lambda$ is the X-ray wavelength, $\theta$ is half of the diffraction angle 2θ, and $h$, $k$, and $l$ are the Miller indices of the lattice plane. The lattice parameter was fitted by an exponential decay function using the Curve Fitting Tool on Matlab using the following formula,

$$a(t) = A * \exp\left(-\frac{t}{\tau}\right) + a_0$$

Where $t$ is time, $A$ is the initial amplitude, $\tau$ is a rate constant, and $a_0$ is the lattice parameter at initial solidification. The function was then used to calculate the cooling rate $\dot{T}(t)$ using the following equation:

$$\dot{T}(t) = \frac{\dot{a}(t)}{CTE * a_0}$$

**Supplementary Material:**

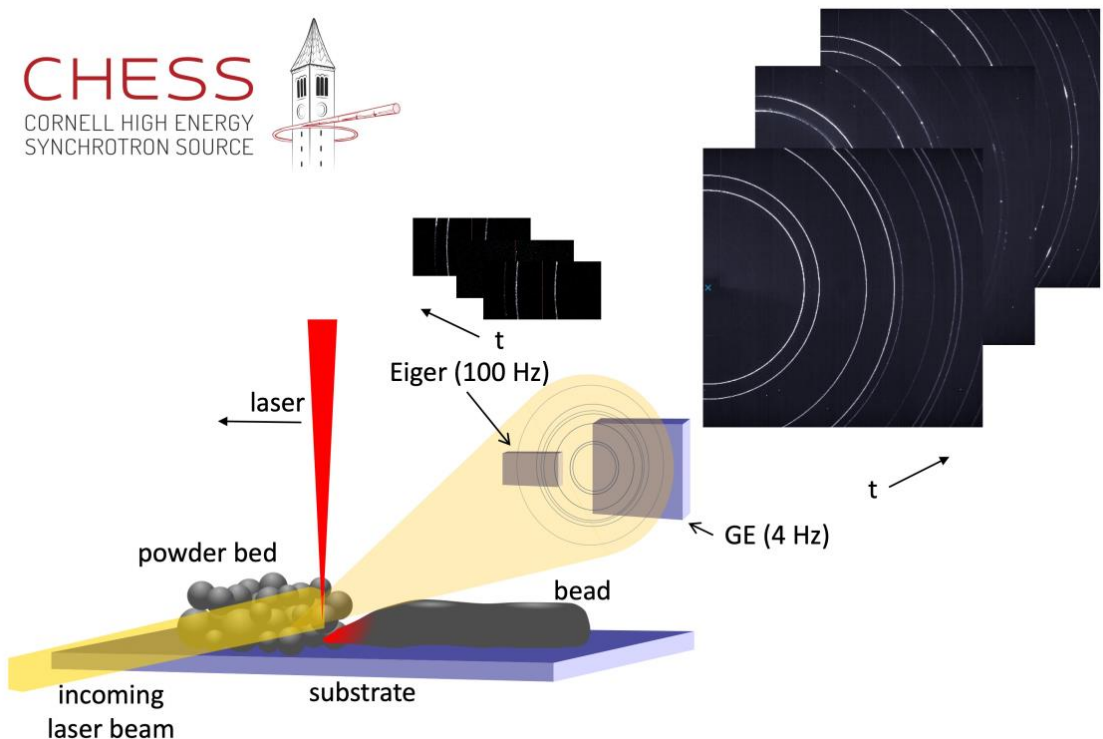

**Fig. S1. Schematic representation of the custom setup integrated at CHESS.** An enclosure (not pictured) maintains an inert atmosphere while the laser beam rasters over a powder bed to simulate an AM environment. Although the powder preparation follows a powder-bed style setup, the scanning speed used in the study results in cooling rates characteristic of a DED processing condition. The transmitted XRD pattern is captured by two detectors (CdTe Eiger 500k area detector and a GE 41-RT area detector) at 100 Hz and 4 Hz, respectively.

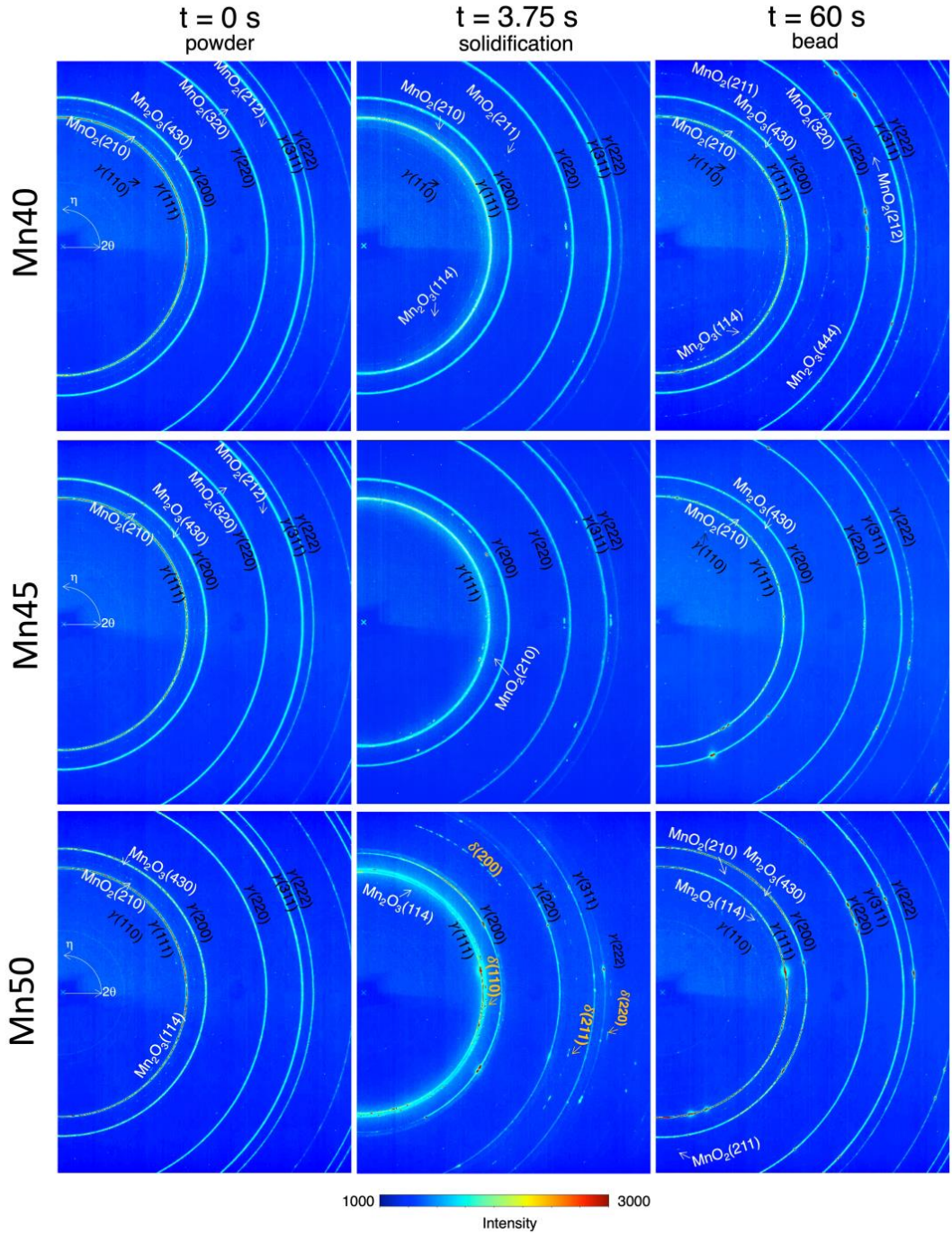

**Fig. S2. GE detector images of XRD patterns in powder, solidification, and bead for Mn40, Mn45, and Mn50.** All rings corresponding to fcc (γ) are annotated in black, those corresponding to bcc (δ) in yellow, and manganese oxides ($MnO_2$ and $Mn_2O_3$) in white. In Mn40 and Mn45, diffracted rings with higher intensities all correspond to the fcc phase throughout the process. In Mn50, all continuous rings correspond to the γ phase from powder diffraction (t = 0 s) as well. However, at the beginning of solidification (t = 3.75 s), additional diffraction spots emerge that correspond to the δ phase. These spots quickly disappear, and after cooling (t = 60 s), only the γ-fcc and manganese oxide rings remain.

**Table S1. Chemical composition of powder and as-printed FeMnCoCr**

| | Composition (at.%) | | | | | |
|---|---|---|---|---|---|---|
| | Mn40 | | Mn45 | | Mn50 | |
| | Powder | Print | Powder | Print | Powder | Print |
| Fe | 40.6 ± 0.3 | 39.5±1.5 | 36.7 ± 1.1 | 36.1±0.9 | 31.7 ± 0.4 | 31.1±1.5 |
| Mn | 40.0 ± 0.2 | 41.2±1.3 | 43.6 ± 1.7 | 45.3±0.6 | 48.8 ± 0.2 | 49.9±1.1 |
| Co | 10.0 ± 0.2 | 9.9±0.4 | 10.1 ± 0.2 | 9.2±0.2 | 10.0 ± 0.1 | 9.8±0.5 |
| Cr | 9.4 ± 0.3 | 9.0±0.4 | 9.6 ± 0.5 | 9.4±0.5 | 9.6 ± 0.2 | 9.2±1.0 |

**Acknowledgments:** A.W. thanks Teri Juarez and Bryan McEnerney for discussions and insights at Jet Propulsion Laboratory and beyond. A.W. and A.M. acknowledge the contributions of Katherine Shanks and Amlan Das during experiments at CHESS and on the data analysis. A.W. and J.B. thank Chenxi Tian, Jinyeon Kim, Dasol Yoon and Claire Matthews for assistance with sample preparation.

**Funding:**

Department of Energy, Office of Science, Basic Energy Science Early Career Award [#DE-SC0022860] (Synchrotron data analysis and microscopy)

National Science Foundation CAREER Award [CMMI-2046523]

NASA Space Technology Graduate Research Opportunities [Grant #80NSSC20K1199]

Cornell Center of Materials Research Shared Facilities, which are supported through the NSF MRSEC program [DMR-1719875]

National Science Foundation [DMR-1829070]

National Science Foundation CAREER Award [CMMI 2047218]

NASA Space Technology Graduate Research Opportunities [Grant # 80NSSC19K1142]

**Author contributions:**

Conceptualization: AW, AM

Methodology: AW, JB, NS, JS, WX, TMS, AM

Investigation: AW, JB, NS, WX, TMS, AM

Visualization: AW, NS, WX, TMS, AM

Funding acquisition: AW, NS, WX, AM

Project administration: AW

Supervision: AM

Writing – original draft: AW

Writing – review & editing: AW, NS, JS, WX, TMS, AM

**Competing interests:** Authors declare that they have no competing interests.

**Data and materials availability:** All data are available in the main text or the supplementary materials.